\documentclass[a4paper,pra,twocolumn]{revtex4-1}  %

\usepackage{graphicx}  
\usepackage{dcolumn}   
\usepackage{bm}        
\usepackage{amssymb}   
\usepackage{hyperref}

\usepackage{soul,color}

\usepackage{caption} 
\usepackage{subcaption} 

\usepackage{amsfonts}

\usepackage{amsmath}
\usepackage{blkarray}
\usepackage{braket}
\usepackage{multirow}
\usepackage{mathtools}

\usepackage{tikz}
\usepackage[utf8]{inputenc}

\newcommand{\w}{\omega}

\newcommand{\sig}{\sigma}

\newcommand{\wx}{\epsilon}
\newcommand{\wrr}{\omega_R}

\newcommand{\N}{\mathcal{N}}
\newcommand{\nb}{N}

\newcommand{\Nd}{\mathcal{N}_D}

\newcommand{\stt}{\hat{\mathcal{S}}}
\newcommand{\stf}{\hat{\tilde{\mathcal{S}}}}

\newcommand{\sttt}{\mathcal{S}}

\newcommand{\sket}[1]{\ket{\mathcal{S}_0(#1)}}
\newcommand{\aket}{\ket{1_P}}
\newcommand{\dsket}{\ket{{\Phi}(\w)}}
\newcommand{\dskett}{\ket{{\Phi}(\tilde\w)}}

\newcommand{\dsketlp}{\ket{{\Phi}(\tilde\w_{LP})}}
\newcommand{\dsketup}{\ket{{\Phi}(\tilde\w_{UP})}}

\newcommand{\wt}{{\tilde\omega}}

\newcommand{\evket}{\ket{{\Psi}(\w)}}
\newcommand{\evkett}{\ket{{\Psi}(\tilde\w)}}

\newcommand{\ev}{\hat{\Psi}^+(\w)}

\newcommand{\nex}{\hat{\mathcal{N}}_{ex}}

\newcommand{\h}{\mathcal{\hat H}}

\newcommand{\hex}{\mathcal{\hat H}_{B,1_{ex}}}

\newcommand{\com}[1]{}

\newcommand{\ad}{\hat{a}^\dagger}
\newcommand{\an}{\hat{a}^{}}
\newcommand{\sd}{\hat{\sigma}^{+}_i}
\newcommand{\sn}{\hat{\sigma}^{-}_i}

\newcommand{\rev}[1]{{\color{blue}{#1}}}  
\newcommand{\az}[1]{{\color{magenta}{#1}}} 

\DeclareMathOperator\erf{erf}
\DeclareMathOperator\erfi{erfi}

\begin{document}


\title{Analytical solution of the disordered Tavis-Cummings model and its Fano resonances}

\author{M. Ahsan Zeb}
\affiliation{Department of Physics, Quaid-i-Azam University, Islamabad 45320, Pakistan}

\date{\today}

\begin{abstract}

$\N$ emitters collectively coupled to a quantised cavity mode are described by
the Tavis-Cummings model.
We present complete analytical solution of the model in the presence of inhomogeneous couplings and energetic disorder.
We derive the exact expressions for the bright and the dark sectors
that decouple the disordered model and find that, in the thermodynamic limit, the energetic disorder transforms the bright sector to
Fano's model 
that can be easily solved.
We thoroughly explore the effects of energetic disorder 
assuming a Gaussian distribution of emitter transition energies.
We compare the Fano resonances in optical absorption and inelastic electron scattering both in the weak and the strong coupling regimes.
We study the evolution of the optical absorption with an increase in the disorder strength
and find that it 
changes the lower and upper polaritons to their broadened resonances that finally 
transform to a single resonance at the bare cavity photon energy,
thus taking the system from the strong to the weak coupling regime.
Interestingly, we learn that 
the Rabi splitting can exist even in the weak coupling regime
while
the polaritonic peaks in the strong coupling regime can represent almost excitonic states at intermediate disorder strengths.
We also calculate the photon Green's function to see 
the effect of cavity leakage and non-radiative emitter losses
and find that 
the polariton linewidth exhibits a minimum as a function of detuning
when the cavity leakage is comparable to the Fano broadening.

\end{abstract}
\maketitle

\com{
1. kappa gamma removed: Fig captions,,,, kappa gamma removed... related paragraphs of fig description... \\
2. read the green sec carefully to see if it is consistent with fano sec... \\
3. terminology polariton and polarito's fano resonance, photons fano resonance...\\
4. referee's recommended discussion and comparison with Herrera and spano etc... ?? cite them and add a short paragraph...\\
5. spelling check... typos... etc.... careful proof reading....
}


\section{Introduction}

Cavity quantum electrodynamics (cavity QED)
deals with the interaction of spatially confined quantised 
electromagnetic field and some form of matter excitations
that can originate from a number of different systems, e.g.,
atoms~\cite{WaltherIOP2006,MivehvarAP21}, 
molecules~\cite{lidzey98,Keeling2020:review}, 
quantum dots~\cite{KirazIOP03,StockklauserPRX17,NajerNat2019}
and Bose-Einstein condensates~\cite{ColombeNat2007,Brennecke2007}. 
In microcavities, 
interesting many-body phenomena can arise
as
the photons mediate interaction~\cite{VaidyaPRX18,GuoPRL19,PeriwalNat2021} and coherence~\cite{GeorgesPRL18} between the matter constituents and in turn acquire correlations among themselves even when they leak out~\cite{ClarkNat20}.
In the strong coupling regime~\cite{DovzhenkoReview18},
these systems host polaritons~\cite{Basov21} that 
have a hybrid matter-light character
and, due to their
technological prospects~\cite{Kasprzak2006,Houck2012,Sanvitto2016,Keeling2020:review,Kravtsov2020,Moxley2021},
have been extensively studied.

The minimal model to describe the cavity QED
with multiple matter systems, or emitters as they are usually called,
is the Tavis-Cummings (TC) model~\cite{TCM68}.
The model
can be analytically solved
in the absence of disorder
using the bright (symmetric) and dark (non-symmetric) states~\cite{Ribeiro2018,SM}.
However, in the presence of 
a disorder in the transition energies of the emitters,
these states are not decoupled
so this transformation is rendered useless and we resort to brute force numerical solutions~\cite{Cwik16,BotzungPRB20,DuPRL22}.
In some cases,
the disorder can even invalidate the basic TC model and 
an extension is deemed necessary~\cite{BlahaPRA2022}.
In natural systems, on the one hand,
disorder both in 
the transition energies 
and couplings 
is inevitable.
On the other, 
a tuneable disorder
can be realised in artificial systems~\cite{HuangACSP19,MazhorinPRA22}.
Thus, exploring the  
effects of disorder is important and 
solving the TC model analytically
can be a significant step forward as it
can provide us with an insight into
the disordered cavity QED systems.

Although the TC model describes a wide range of systems,
we illustrate our results here considering organic microcavities
as an example.
The transition 
energies of organic molecules 
usually have a 
distribution of width ${\sqrt{2}\times 0.1eV}$~\cite{bassler81, bassler82,fnote},
which is a sizeable fraction of
the typical Rabi splitting of $0.5eV$~\cite{TropfAOM18}
in these systems.
Furthermore,
the distribution of orientations and positions of the molecules in the cavity translates to a disorder in their couplings to the cavity mode.

Organic polaritons have interesting properties, e.g.,
condensation and lasing at room temperature~\cite{Kena-Cohen10,Plumhof14,Keeling2020:review}.
Their effect on 
materials' properties, e.g.,
charge and energy transport~\cite{Feist2015, Hagenmueller2017,Schafer2019,zeb2020}, and
chemical reaction rates~\cite{Thomas2019,HerreraPRL16,Galego2016,galego17,MartinezACS2018,ebbesen2016hybrid,feist2017polaritonic,Ribeiro2018}
is also studied recently.
Besides the polaritons,
we have a large number of
dark states in such systems~\cite{Ribeiro2018,KenaCohenACSCS19},
that also play important role in various 
processes of interest, e.g., catalysis~\cite{DuPRL22}.
While the dark states were initially thought to be only a reservoir of 
incoherent excitations~\cite{Agranovich2002,Agranovich2003},
they are now well appreciated for their coherence
and delocalised nature~\cite{GonzalezPRL16,HuJCP18,PandyaAS22}.
Although, the dark states do not couple to the light,
they can still be excited optically indirectly by exciting polaritons or higher energy emitter states.
In addition, the 
electrical excitation predominantly creates the dark states
due to their large density of states.
The dark states can relax to the lower polariton state,
which is important for creating large polariton population
 for condensation and lasing for instance,
and their dynamics has been studied extensively~\cite{pino2018,groenhof2019,sommer2021,Eizner2019,polak20,yu2021,mewes2020,wersal2019,xiang2019}.
The effect of disorder on the localisation of the dark state has also been numerically studied recently~\cite{BotzungPRB20}.
Due to their hugely important role in organic microcavities and other cavity QED systems,
it is desired to know their exact form under realistic conditions, 
which can lead to a significant development of analytical methods.


Since real systems are often weakly coupled to their environment,
the resulting \emph{homogeneous} broadening 
of the cavity and molecular states
is inherited 
by the polariton states, and can be treated 
using open quantum system approaches---master equations---or 
phenomenologically with the Green's functions
\cite{ThompsonPRL92,HerreraPRA17,ZhangJPB21}
(or an equivalent quantum control method~\cite{DongIEEE22}).
In fact, 
to obtain the optical absorption spectrum,
the Green's functions can 
exactly treat the energetic disorder or the 
\emph{inhomogeneous} broadening as well~\cite{MazhorinPRA22,GeraJCP22},
as we will later show in this article.
(Nevertheless, it has also been used to find 
an approximate solution by neglecting 
the effect of energetic disorder on the light-matter coupling strength~\cite{Cwik16}.)
However, only the photon Green's function can be computed this way,
which alone
cannot describe the eigenstates of the system, 
thus severely limiting the scope of the previous
studies~\cite{Cwik16,MazhorinPRA22,GeraJCP22,DongIEEE22}.

The exact numerical diagonalisation, 
another method that is simple and usually effective,
has its own shortcoming.
In a clean system with identical emitters,
studying the collective effects in the
thermodynamic limit
is possible
as the typical size that is considered large enough 
is $\N=20$, where $\N$ is the number of emitters.
However, in an energetically disordered system,
a realistically smooth distribution of states
that converges the results 
requires $\N \gg 10^6$,
which is computationally intractable
considering that \emph{all} eigenstates of the Hamiltonian 
are to be computed. 
This is the reason that exact diagonalisation 
in such cases would even
fail to capture the features in the optical spectrum 
that are given
by the photon Green's function
in the thermodynamic limit.

\com{
ref[23,25]: (Cwik16, Ribeiro 18)

--- Ribeiro 18: do not consider energetic disorder (but analyse the effect of a band of cavity modes to some extent).
--- Cwik16: consider energetic disorder but only approximately using greens functions... equiv to flat V.; 
--- GeraJCP22: treat energetic disorder exactly using greens functions... but do not consider the bright-dark decoupling of the Hamiltonian --- which is essential to obtain the Fano's model. 
\rev{
energetic disorder/inhomogeneous:\\
--- cwik16: (temperature and diamagnetic term) small energetic disorder ($\sigma<\kappa$, see Fig.1 caption) approximate fashion in the green's functions, ignoring its effect on the coupling --- equivalent to a considering flat emitter DOS;
--- shoaib; vibr dirodered sys Wajahat wala paper... \\
--- Agronovich prb2003 paper, classical approach for inhomogeneous broadening...\\
losses or homogeneous broadening:\\
various approaches.... 
---Green's functions (standard textbook level)\\
---mapping to linear systems (Dong IEEE),\\
works:\\
--- Kimble 92(ThompsonPRL92), Herrera and spano PRA17,  ZhangJPB21, Ribeiro 18, cwik16
}

4. emphasise the difference between homo and inhomo... in our case...??\\
}

\com{Fano resonances that 
arise when a discrete or localised state weakly interacts with a continuous band~\cite{Fano61,MiroRMP10,KamenetskiiBook18,Cao20Review}.
}

What is so important in the thermodynamic limit
of the energetically disordered system that has not yet been 
understood?
It is the emergence of the Fano's model~\cite{Fano61},
which can only be found if 
the Hamiltonian is first decoupled into its exact bright and dark subspaces.
Fano's model describes the interaction of a discrete or localised state to a continuous band and exhibits Fano resonances~\cite{Fano61,MiroRMP10,KamenetskiiBook18,Cao20Review,ZebFano}.
It gives a characteristic asymmetric lineshape in excitation spectra
that finds applications~\cite{Limonov21} in sensing and switching devices~\cite{Nozaki13,Stern2014}. 
It is ubiquitous in physical systems
~\cite{ZhangPRL06,ZhangPRL08,Tang2010,Ott2013,LuoPRL14,ZielinskiPRL15,WangPRL17,Limonov2017}
 but, being a coherent phenomenon, 
 finding it in a disordered system~\cite{Poddubny2012,Silevitch2019} is not common.
The Fano's model has always been applied in the weak coupling regime
but, as we will shortly see, it is also relevant in the strong coupling regime~\cite{ZebFano}.

\begin{figure}
\centering
\includegraphics[width=1\linewidth]{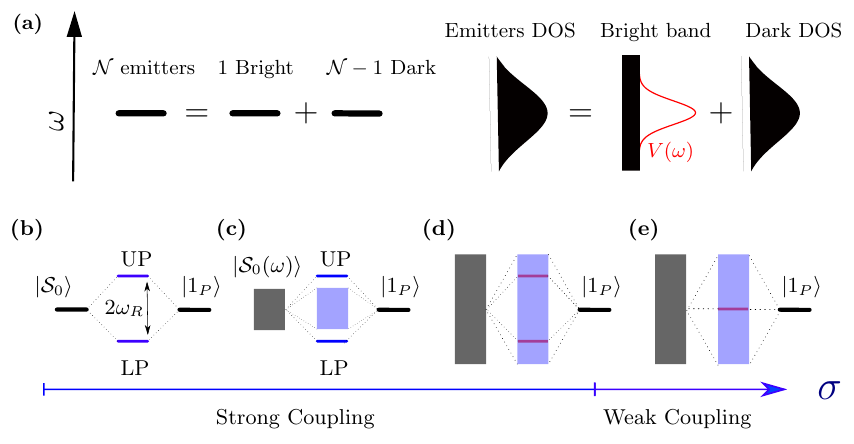}
\caption{ Effect of energetic disorder in the thermodynamic limit.
(a) Bright (permutation symmetric) and dark (non-symmetric) states. 
 In the absence of energetic disorder (left),
 $\N$ emitters form a single bright state and ${\N-1}$ dark states.
 In the presence of energetic disorder (right),
 where the density of states (DOS)
 forms a localised distribution,
 such bright-dark decoupling occurs at all energies \emph{separately}, 
 which
 produces a non-degenerate continuum of bright states (flat DOS)
 and a distribution of dark states. 
 The coupling of the bright band to the cavity mode $V(\w)$ 
 equals the square root of the emitter DOS.
(b-e)
The eigenstates of the system 
formed by the coupling between
the bright state ${\ket{\mathcal{S}_0}}$ or band ${\ket{\mathcal{S}_0(\w)}}$ and the cavity photon state ${\ket{1_P}}$ .
Without energetic disorder (b), 
upper and lower polaritons (UP, LP) are formed.
With energetic disorder (c-e),
we have a continuum of eigenstates
at the original bright band energies
but the true polariton states exist only at small 
disorder (c).
At intermediate disorder,
we have polariton's Fano resonances (d)
that transform to the photon's Fano resonance (e)
at large disorder strengths.  
}
\label{fig:cartoon}
\end{figure}

Here, we consider the Tavis-Cummings model with disorder both in emitter energies and their couplings
to the cavity mode.
We 
 derive the exact expressions for the bright and the dark states that 
span the two decoupled subspaces of the model.
We show that,
in the thermodynamic limit,
the bright states of the disordered system
 form a non-degenerate \emph{band} whose interaction with the 
cavity mode gives the Fano's model
that can be exactly solved.
Figure~\ref{fig:cartoon} illustrates the
formation of the bright band
and the nature of the eigenstates of the system
at varying levels of the disorder strength.
In Fig.~\ref{fig:cartoon}(a),
when $\N$ emitters have the same transition energy, 
their states can be transformed to give a single bright state ${\ket{\mathcal{S}_0}}$ 
that couples to the cavity
and ${\N-1}$ dark states that are decoupled, hence the names.
However, when the emitter energies form a distribution,
a similar transformation at every energy
produces bright and dark states at that energy,
decoupling the dark sector completely
and forming a non-degenerate band of bright states ${\ket{\mathcal{S}_0(\w)}}$ with flat density of states (DOS).
The eigenstates of the system in the single excitation space are
formed by the coupling between
${\ket{\mathcal{S}_0}}$ or ${\ket{\mathcal{S}_0(\w)}}$ and the cavity photon state ${\ket{1_P}}$
as shown in Fig.~\ref{fig:cartoon}(b-e),
where different scenarios that give polaritons [Fig.~\ref{fig:cartoon}(b,c)],
their Fano resonances [Fig.~\ref{fig:cartoon}(d)]
and the photon's Fano resonance [Fig.~\ref{fig:cartoon}(e)] are illustrated,
which result from progressively increasing the energetic disorder $\sigma$
that takes the system from the strong coupling regime to the weak coupling regime.
Our calculations suggest that
the polaritons are a manifestation of the Fano resonances at weak disorder.
Further, we show that the Rabi splitting or the anticrossing
in the optical spectra does not have to correspond to the polaritons, they should also be seen when the corresponding Fano resonances are better thought of as
matter states with strong optical absorption.

For completeness, 
we also consider the effect of losses on the the optical response of the system and
compute the exact photon Green's function
that encodes it.
We explore the effect of cavity and emitter losses
on the optical absorption spectrum
and find that, in organic microcavities,
it should exhibit a minimum in the polariton linewidth
around zero detuning. 
While preparing for the revised version of this manuscript,
we found another work with some overlap~\cite{GeraJCP22}.

The organisation of this article is as follows.
The decoupling of the TC model into its bright and dark spaces
is presented in sec.~\ref{sec:model},
where sec.~\ref{sec:identical} considers $\N$ identical emitters,
sec.~\ref{sec:offdiag} includes disorder in the couplings only,
while sec.~\ref{sec:fully} also takes account of energetic disorder.
Section~\ref{sec:emerge}
contains the results related to the Fano's model.
The emergence of the Fano's model
and its solution
 in two possible scenarios [shown in Fig.~\ref{fig:cartoon}(c,d)]
are given in sec.~\ref{sec:fano}.
Fano resonances in two types of excitation spectra---optical 
absorption and inelastic electron scattering---are
 presented in sec.~\ref{sec:fanosol} both in the weak and strong coupling regimes.
Section~\ref{sec:absorption} explores the effects of energetic disorder,
where sec.~\ref{sec:rabi} 
considers the Rabi splitting 
and 
the nature of the eigenstates,
sec.~\ref{sec:homo}
discusses the differences from 
a homogeneous broadening, 
and 
sec.~\ref{sec:fbroad}
analyse the Fano 
broadening of polaritons.
Finally,
the effects of losses are described in sec.~\ref{sec:greenlosses},
where the photon Green's function is used to study
the polariton lineshape and linewidth
as well as the role of the distribution of emitters' energies.

\com{Expressions for two types of excitation spectra ---
optical absorption and inelastic electron scattering ---
are discussed in sec.~\ref{sec:avsm},
and Fano resonances in these spectra
are compared both in the weak and strong coupling regimes
in sec.~\ref{sec:fres}.
}

\section{Model and its solution}
\label{sec:model}

Consider $\N$ two-level emitters coupled to a common cavity mode.
The Hamiltonian of this system in the rotating wave approximation~\cite{zubairy1997} is given by
\begin{multline}
\label{eq:h}
\h
= 
\w_c \ad\an
   + 
  \sum_{i=1}^\N \Big \{ \wx_{i}  \hat \sig_i^+ \hat \sig_i^-
  +  g_i(\ad \sn + \sd \an)\Big\},
 \end{multline}
where $\w_c$ is the energy of a cavity photon,
$\ad,\an$ are its creation and annihilation operators,
and $\sig_i^{\pm}$ are raising and lowering operators for the $i$th emitter
that has transition energy $\wx_i$ and couples to the cavity mode with strength $g_i$.
Both $g_i$ and $\wx_i$ are
\emph{randomly} distributed according to some probability functions.
$\h$ commutes with the excitation number ${\nex=\ad\an + \sum_i^\N \sd\sn}$,
which allows its diagonalisation in an eigenspace of $\nex$.
We will focus on the single excitation subspace in this article.

In the following sections, 
we describe the decoupling of the emitters' Hilbert space into 
bright and dark sectors, where the bright sector couples to 
the cavity mode but the dark sector does not. 

\subsection{Identical Emitters}
\label{sec:identical}

For identical emitters, i.e., when ${\wx_i=\wx}$ and ${g_i=g}$ for all $i$,
the solution is already well known.
In this case,
a unitary transformation to 
the symmetric (bright) and non-symmetric (dark) superpositions of the emitter states block-diagonalises the TC model in the single excitation space.
The bright state is created by
\begin{eqnarray}
\stt^+_0 &\equiv& \frac{1}{\sqrt{\N}}\sum_{i=1}^\N\hat\sig_i^+,
\end{eqnarray}
and produces the two polariton states due to its coupling to the cavity mode,
which are given by,
\begin{align}
\label{eq:2ls}
\hat\Psi_{UP}^{+} &= \cos\theta \ad + \sin\theta \stt^+_0,\\
\hat\Psi_{LP}^{+} &= \sin\theta \ad - \cos\theta \stt^+_0,\\
\label{eq:EUP}
\w_{UP} &= \frac{1}{2}(-\delta + \sqrt{\delta^2+4\wrr^2}),\\
\label{eq:ELP}
\w_{LP} &= -\frac{1}{2}(\delta + \sqrt{\delta^2+4\wrr^2}),\\
2\theta &= \tan^{-1}(2\wrr/\delta),
\end{align}
where $\delta=\w_c-\epsilon$ and
 $\hat\Psi_{UP/LP}^{+}$ create the polaritonic states
at energies $\w_{UP/LP}$.

For the non-symmetric or dark states, being a degenerate manifold (at the bare exciton energy $\wx$), there is no unique representation and one can choose any complete set as basis states for the dark space.
Usually, a set of delocalised
discrete Fourier transform states is used, given by
\begin{eqnarray*}
\stf^+_{k} &\equiv& \frac{1}{\sqrt{\N}} \sum_{n=1}^{\N} e^{i 2\pi k n/\N}\hat\sig_{n}^+,~ k\in[1,\N-1].
\end{eqnarray*}
In case of inhomogeneous couplings,
(i.e., when a disorder in the couplings exists), 
these states no longer represent the dark sector.
Here, we present another set 
that can be easily modified in such a case. 

Inspired from the maximally localised dark state,
\begin{eqnarray*}
\stt^+_{\N-1} &\equiv& \sqrt{\frac{\N-1}{\N}}\hat\sig_{\N}^+
-\frac{1}{\sqrt{\N(\N-1)}} 
\sum_{i=1}^{\N-1} \hat\sig_i^+,
\end{eqnarray*}
that is fairly known and is localised at site $\N$,
we can focus on how it turns out to be orthogonal to the bright state $\stt^+_0$.
We note that
the contribution from the $\N$th site is exactly cancelled by the sites $1$ to $\N-1$.
Designing another dark state on the same pattern 
to ensure its orthogonalisation to $\stt^+_{\N-1}$,
we obtain $\stt^+_{\N-2}$ that 
is maximally localised on emitter $\N-1$ but
has zero component along $\hat\sig_\N^+$, i.e., on the site $\N$.
Repeating this process, we construct an orthonormal representation for the dark states
where different dark states tend to localise on different molecules, 
albeit to a varying degree, 
given by~\cite{zebPRA21}
\begin{eqnarray}
\stt^+_j &\equiv& \frac{1}{\sqrt{j(j+1)}} 
\left( \sum_{i=1}^j \hat\sig_i^+ - j \hat\sig_{j+1}^+  \right),
\end{eqnarray}
where ${j\in[1,\N-1]}$.

\subsection{Only off-diagonal disorder}
\label{sec:offdiag}

It is well known that when a disorder in the couplings exists,
$\stt^+_0$ modifies to  
\begin{eqnarray}
\label{eq:s0}
\stt^+_0 &\equiv& \frac{1}{G_\N}\sum_{i=1}^\N g_i\hat\sig_i^+,
\end{eqnarray}
where 
${G_\N^2 \equiv \sum_{i=1}^\N g_i^2}$.
However, exact expressions for 
the dark states are not known. 
As we mentioned in the previous section, we modify the maximally localised states 
to obtain just that.
These can be obtained by following the changes in $\stt^+_0$
and imposing the orthogonalisation of the dark states to this bright state.
That is, try
\begin{eqnarray*}
\stt^+_j &\equiv&
\alpha_j\left( \sum_{i=1}^j g_i\hat\sig_i^+ - \beta_j \hat\sig_{j+1}^+  \right),
\end{eqnarray*}
and impose $\braket{GS|\stt^-_0\stt^+_j|GS}=0$,
where $\ket{GS}$ is the ground state of the emitters.
This gives,
${\beta_j=G_j^2/g_{j+1}}$,
and then normalisation 
${\braket{GS|\stt^-_j\stt^+_j|GS}=1}$
gives ${\alpha_j=g_{j+1}/(G_jG_{j+1})}$.
We thus obtain,
\begin{align}
\label{eq:sj}
\stt^+_j &\equiv \frac{g_{j+1}}{G_jG_{j+1}} 
\sum_{i=1}^j g_i\hat\sig_i^+
-\frac{G_j}{G_{j+1}} \hat\sig_{j+1}^+.
\end{align}

\subsection{Fully disordered}
\label{sec:fully}

Due to the presence of disorder in the diagonal terms in $\h$, 
the above transformations 
no longer work.
However, as we show below, 
the \emph{exact} bright and dark subspaces still do exist
and it is still possible to block diagonalise $\h$.
Even though the energies $\{\wx_i\}$ are randomly distributed, 
assuming their intrinsic linewidth $\gamma$ to be negligible for the moment,
we can count the number of states at any given energy
and make bright and dark sectors for every energy level.
Indexing these energy ``levels'' in ascending order,
let's take $K_n$ to be the degeneracy of the $n$th level at energy $\w_n$.
Relabelling the emitters ${i\to n,i_n}$
with $i_n\in[1,K_n]$
and
dropping the subscript for brevity, 
${\wx_{i}\to\w_{n,i}}$, 
${g_{i}\to g_{n,i}}$, and
${\sig^\pm_{i}\to\sig^\pm_{n,i}}$.
This 
allows us to make the bright and the dark states for every level,
similar to Eqs.~\ref{eq:s0}-\ref{eq:sj}.
For $n$th degenerate level, 
we have
\begin{align}
\label{eq:sn0}
\stt^+_{n,0} &\equiv \frac{1}{G_{n,K_n}}\sum_{i=1}^{K_n} g_{n,i}\hat\sig_{n,i}^+,\\
\label{eq:snj}
\stt^+_{n,j} &\equiv \frac{g_{n,j+1}}{G_{n,j}G_{n,j+1}} 
\sum_{i=1}^j g_{n,i}\hat\sig_{n,i}^+
-\frac{G_{n,j}}{G_{n,j+1}} \hat\sig_{n,j+1}^+, 
\end{align}
where ${j \in [1,K_n-1]}$ and ${G_{n,l}^2 = \sum_{i=1}^l g_{n,i}^2}$.
For the non-degenerate levels, 
only a bright state $\stt^+_{n,0}=\hat\sig_{n,1}^+$ exists.
Taking
$\nb$ as the number of the bright states or the number of levels,
$\h$ can be written as ${\h=\h_B + \h_D}$ where
\begin{multline}
\label{eq:hb}
\h_B = \w_c\ad\an + \sum_{n=1}^{\nb} \w_n \stt^+_{n,0}\stt^-_{n,0}  \\
+ \sum_{n=1}^{\nb} 
G_{n,K_n} \left[ \an \stt^+_{n,0} + \ad \stt^-_{n,0}  \right],
\end{multline}
\begin{flalign}
\label{eq:hd}
~~~\h_D &= \sum_{n=1}^{\nb}  \w_n \sum_{j=1}^{K_n-1}\stt^+_{n,j}\stt^-_{n,j},&&
\end{flalign}
where
$\h_D$ is again already diagonal,
and 
there are ${\Nd=\sum_{n=1}^{\nb} (K_n-1)=\N-\nb}$ dark states in total.
Thus, to completely solve 
the disordered TC model,
all that is left is 
the diagonalisation of $\h_B$ above.

$\h_B$
can be diagonalised numerically. 
The size of the bright space could be negligibly small compared to the full space, i.e.,
$\nb\lll \N$. For instance, even at $\N=10^{23}$, $\nb$ could be taken as small as ${\sim 10^2-10^3}$ for most realistic situations without compromising the thermodynamic limit.
However, probably the best outcome of our approach is that,
at large $\N$, 
when $\{\w_n\}$ forms a continuous band,
following Fano~\cite{Fano61},
analytical solution of $\h_B$ can also be found, as follows.

\com{
$\h_B$
can be diagonalised numerically. 
The size of the bright space could be negligibly small compared to the full space, i.e.,
$\nb\lll \N$. For instance, even at $\N=10^{23}$, $\nb$ could be taken as small as ${\sim 10^2-10^3}$ for most realistic situations.
In a clean system with identical emitters, 
$\N=20$ is the typical size that is considered large enough for 
studying the collective effects in the
thermodynamic limit.
However, in an energetically disordered system,
numerically solving the full Hamiltonian $\h$ --- without decoupling it into the bright and dark sectors ---
requires
$\N$ to be large enough ($\gg 10^6$) to give 
a realistically smooth distribution of states.
Computing \emph{all} eigenstates of $\h$ 
is computationally intractable for such large systems,
as even at ${\N=10^6}$ the memory required just to store the states 
is ${64}$TBs for double precision real numbers.
This is the reason that previous numerical studies
that consider energetic disorder
~\cite{Cwik16,zebPRA21}
never reached the conclusions 
that our approach finds.
However, probably the best outcome of our approach is that,
at large $\N$, 
when $\{\w_n\}$ forms a continuous band,
following Fano~\cite{Fano61},
analytical solution of $\h_B$ can also be found, as follows.
}

\section{Emergence of Fano's model} 
\label{sec:emerge}


In the thermodynamic limit, $\N\to\infty$,
we can consider the continuous limit for the bright band.
Taking 
${G_{n,K_n}^2 \to  V(\w)^2 d\w }$,
\begin{align}
\nonumber
V(\w)^2 d\w &= K(\w)d\w \int dg P_c(g) g^2,\\ 
\label{eq:Vw0}
&=  \N P(\w) d\w \braket{g^2},
\end{align}
where $P_c(g)$
and $P(\w)$ are distribution functions for couplings and energies.
Note that
as long as we get the same value for
$\braket{g^2}$, the specific form of
$P_c(g)$ does not matter.
Taking ${\wrr^2 \equiv \sum_n G_{n,K_n}^2}$
and noting that $\int{P(\w) d\w}=1$, 
we can now write, 
$\wrr^2 = \int{V(\w)^2 d\w} = \N \braket{g^2}$,
which can be compared back to Eq.~\ref{eq:Vw0} above
to write
\begin{align}
\label{eq:Vw}
V(\w) = \wrr \sqrt{P(\w)}.
\end{align}

Thus, the continuous limit of 
$\h_B$ in Eq.~\ref{eq:hb} becomes,
\begin{multline}
\label{eq:hbw}
\h_B = \w_c\ad\an + \int  \w \stt^+_0(\w)\stt^-_0(\w) d\w \\
+ \int 
V(\w) \left[ \an \stt^+_0(\w) + \ad \stt^-_0(\w) \right] d\w.
\end{multline}

\subsection{Fano's model: $\h_B$ in ${N_{ex}=1}$ subspace}
\label{sec:fano}

There are three ingredients in the
Fano's model~\cite{Fano61}: 
(i) a discrete state interacting with a 
(ii) continuum of states
that form the eigenstates of the system
to which (iii) transition from another discrete state 
[uncoupled to (i) and (ii)]
 is to be explored.
Since $N_{ex}$ is a conserved quantiy,
$\h_B$ is
decoupled 
in subspaces with different $N_{ex}$.
In the following we show that $\hex$ ---
$\h_B$ 
in $N_{ex}=1$ subspace ---
consists of (i) and (ii), whereas the transitions from the 
sole eigenstate of $\h_{B,0_{ex}}$
(the ground state of the full system)
to the eigenstates of $\hex$
should give the Fano resonances.

In the single excitation subspace
we can either have a cavity photon 
or a molecular excitation of the bright band.
The corresponding states are ${\ket{1_P}\equiv\ad\ket{vac}}$
and
${\ket{\sttt_0(\w)}\equiv\stt^+_0(\w)\ket{vac}}$,
where ${\ket{vac}=\ket{0_P}\otimes\ket{GS}}$ is 
the ground state of the full system
that constitutes the subspace with zero excitation ${N_{ex}=0}$.
So projecting $\h_B$ to ${N_{ex}=1}$ subspace
gives
\begin{multline}
\label{eq:hex}
\hex = \w_c\ket{1_P}\bra{1_P} + \int  \w \ket{\sttt_0(\w)}\bra{\sttt_0(\w)}d\w \\
+ \int 
V(\w) \left(\frac{}{} \ket{1_P}\bra{\sttt_0(\w)} + \ket{\sttt_0(\w)}\bra{1_P} \right) d\w.
\end{multline}

We can see that
$\hex$ describes the interaction of a localised state 
${\ket{1_P}}$
with a continuum ${\ket{\sttt_0(\w)}}$.
This model 
has been solved by Fano long ago~\cite{Fano61}
to explain the resonance bearing his name 
that occurs at small $V(\w)$.
Due to its relevance to the disordered TC model as shown above,
we thoroughly explore this model in the strong coupling regime elsewhere~\cite{ZebFano} 
where the continuum bandwidth $W$ is assumed to be the largest energy scale
and all eigenstates lie within it.
In real systems, e.g., organic microcavities, however,
 the distribution $P(\w)$ is non-zero only in a finite energy window, which determines the bandwidth $W$ of the bright states.
So the following two scenarios are possible for the eigenstates of $\hex$
in general.

\com{
Here, we briefly give the expressions for the eigenstates originally derived by Fano~\cite{Fano61},
and present results for two types of excitation spectra,
both in the weak as well as the strong coupling regime.
As the correspondence between the creation operators and 
the state vectors is trivial 
(compare Eq.~\ref{eq:evec} below with Eq.~2 in Refs.~\cite{ZebFano,Fano61}),
we keep using creation operators for consistency.

In real systems with finite bandwidth $W$,
 [$\sim6\sigma$ for a Gaussian distribution $P(\w)$],
}

\subsubsection{Small bandwidth: polaritons outside the continuum}
\label{sec:polout}

At $W\lesssim2\wrr$,
depending on the cavity detuning from the centre of $V(\w)$,
we can have one or two polaritons 
as discrete states
outside the bright band,
along with a continuum of eigenstates at the original bright band energies,
as shown in Fig.~\ref{fig:cartoon}(c).
These eigenstates can be easily calculated, see Appendix~\ref{asec:phipsi}.

Assuming $\dsket$ represents the discrete eigenstates
of $\hex$
outside the bright band at energy $\w=\wt$,
we find that
\begin{align}
\label{eq:dsket}
\dskett = \aket + \int {\frac{V(\w')}{\wt-\w'} \sket{\w'} d\w'},
\end{align}
where 
the integration is over the bright band and
the normalisation 
is left on purpose.
The associated eigenvalues $\wt$
are given by the roots of
\begin{align}
\label{eq:roots}
\wt-\w_c-F(\wt)=0
\end{align}
where 
\begin{align}
\label{eq:fwint}
F(\wt) &= \int{\frac{V(\w')^2}{\wt-\w'} d\w'}.
\end{align}
Note that 
there is no singularity in the above integral
as, by definition, ${\w'\neq\wt}$ in this case.


The roots of Eq.~\ref{eq:roots} can also lie within the bright continuum
where $F(\w)$ would be the principal value ($p.v.$) of the integral in Eq.~\ref{eq:fwint}.
There can be up to three roots in the strong coupling case~\cite{ZebFano}
that can be named after the states they belong to in the extreme cases
of weak and strong coupling regimes by
considering a localised distribution $P(\w)$ 
[remember that ${V(\w)=\wrr \sqrt{P(\w)}}$].
If the width of $P(\w)$ is taken to be $\sigma$, 
then at ${\sigma/\wrr\gg1}$ we will only have a single root 
${\wt_c}$ at (or near) the cavity photon energy,
while in the opposite case ${\sigma/\wrr\ll1}$
two more roots $\wt_{LP}$ and $\wt_{UP}$
exist at the lower and upper polariton energies.
$\wt_{LP}$ and $\wt_{UP}$ can be inside the continuum or outside it.
When inside, 
the corresponding states $\dsketlp$ and $\dsketup$
are still eigenstates of $\hex$
and produce polariton's Fano resonances.
This will be explained in the following 
when we discuss 
the structure of the continuum eigenstates.

The eigenstates of $\hex$ within the bright continuum 
have already been calculated by Fano~\cite{Fano61},
where he correctly handled the singularities involved.
Assuming $\evket$ is such
an eigenstate at energy $\w$,
we can write it as a superposition (Appendix~\ref{asec:phipsi}),
\begin{multline}
\label{eq:evket}
\evket = \frac{\sin\Delta(\w) }{\pi V(\w)} \dsket
- \cos\Delta(\w)  \sket{\w},
\end{multline}
where 
\begin{align}
\label{eq:Dw}
\Delta(\w) &= - \arctan\left[\frac{\pi V(\w)^2}{\w-\w_c-F(\w)}\right].
\end{align}
The significance of the state $\dsket$ can now be appreciated further
by noting the sine and cosine terms in the coefficients
of the two component states in this expression.
For a physical process that can excite the system from $\ket{vac}$
state to the eigenstate $\evket$,
there are two transition paths available---via 
$\dsket$ and $\sket{\w}$ components---that 
can interfere if the corresponding amplitudes are finite for the two.
This becomes quite interesting 
at the roots of Eq.~\ref{eq:roots} $\wt$ inside the continuum,
where the phase ${\Delta(\w)}$ jumps between $\pm\pi/2$
and,
 since sine and cosine functions have opposite parity,
the two paths
interfere constructively on one side of the root 
and destructively on the other side,
leading to an asymmetric lineshape there.
This is the Fano resonance that will be discussed later in sec.~\ref{sec:fanosol}.
Since $\Delta=\pm\pi/2$ at $\wt$,
${\sin\Delta=\pm1, \cos\Delta=0}$ there, 
and ${\evkett=\dskett/\pi V(\wt)}$
showing that $\dskett$ also represents the eigenstates
inside the continuum at ${\wt_c,\wt_{LP},\wt_{UP}}$.
In the weak coupling case, when only a single root $\wt_c$ exists,
it will be called the photon's Fano resonance.
In the strong coupling case, 
if the roots $\wt_{LP}$ and/or $\wt_{UP}$
lie within the continuum,
the resonances will be
reminiscent of the discrete polariton states, 
and will hence be called polariton's Fano resonances.

\com{
Later when we consider the eigenstates lying inside the bright band,
there would be a singularity at $\w'=\w$ and 
the above integral will be defined in the sense of its principal value.
}

\subsubsection{Large bandwidth: polaritons inside the continuum}
\label{sec:polin}

At $W\gtrsim2\wrr$,
we do not have any discrete polariton eigenstates
because
all roots of Eq.~\ref{eq:roots} lie within the bright band,
as shown in Fig.~\ref{fig:cartoon}(d,e).
In this case,
all eigenstates are described by
$\evket$ in Eq.~\ref{eq:evket}.
This completes the solution of the disordered TC model
in the thermodynamic limit for a generic distribution $P(\w)$.

\com{
One root of Eq.~\ref{eq:roots} 
always lies within the bright band
and is reminiscent of the weak coupling case 
(where it is the only possible root)~\cite{ZebFano},
as will be shown later 
in sec.~\ref{sec:xx}
when we compare the spectra in the two coupling regimes.
\az{In the strong coupling case,
Eq.~\ref{eq:roots} has
one or two additional roots that
give the discrete polaritons $\dsket$ when outside the bright band,
but described by 
$\evket$ in Eq.~\ref{eq:evket} otherwise.}
This completes the solution of the disordered TC model
in the thermodynamic limit for a generic interaction $V(\w)$. 
}

An interesting situation arises when
a polaritonic root ${\wt=\wt_{LP/UP}}$
lies inside the bright band
but the coupling ${V(\w)\to 0}$ around its position.
In such a case,
${\Delta(\w)\to 0}$ at ${\w\neq \tilde \w}$,
but ${\Delta(\tilde \w)\to \pm \pi/2}$ still holds.
So, ${\evket\to\sket{\w}}$ at ${\w\neq \tilde \w}$,
and ${\ket{\Psi(\tilde\w)}\propto\ket{\Phi(\tilde\w)}/V(\tilde \w)}$ as always.
This observation can be used to unify the above two cases
of small and large bandwidth into the latter,
by extending the bright band on either side and simultaneously 
reducing the coupling ${V(\w)\to0}$ in the extended region.
Considering that ${V(\w)}$ is sharply localised naturally
as ${V(\w)^2=P(\w)}$ can
be assumed to be
 a Gaussian distribution for energetic disorder in 
real systems, 
the above trick will be used henceforth
and $W$ will be assumed to be the largest energy scale.

Let's take
\begin{align}
\label{eq:Pw}
P(\w)=\frac{1}{\sqrt{2 \pi } \sigma}e^{-\w^2/2 \sigma ^2},
\end{align}
where $\sigma$ is the standard deviation that 
determines the width of the distribution
and the mean energy ${\w_0=\int{\w P(\w)d\w}}$
 is taken as a reference.
We can now evaluate $F(\w)$ in Eq.~\ref{eq:fwint}
to obtain
\begin{align}
\label{eq:fintg}
F(\w) &= \frac{\wrr^2 }{\sqrt{2 \pi }\sigma}
\times p.v. \int {\frac{e^{-\w'^2/2 \sigma ^2}}{\w-\w'}d\w'},\\
\label{eq:fw}
F(\w)  &= \pi \wrr^2 P(\w) \erfi\left(\frac{\w}{\sqrt{2} \sigma }\right),\\ \nonumber
\erfi(\w) &\equiv \erf(i \w)/i,
\end{align}
where $\erfi$ is the imaginary error functions.
The contribution of the tail of the Gaussian 
(that may not be present in real systems)
in the above integral is negligible 
(even around ${\w=\w'}$ on the tail --- note that the $p.v.$ of the integral 
in Eq.~\ref{eq:fintg}
measures 
the asymmetry of $e^{-\w^2/2 \sigma ^2}$ around $\w$ in a \emph{small}
energy window as 
the denominator causes a
suppression
away from $\w$).

The response of the system
to an optical, electronic, or a hybrid excitation
can now be evaluated. 
To illustrate the characteristic nature of 
 the eigenstates,
  let's first compare the
 inelastic electron scattering
 with the optical absorption.
\com{ The bare transition amplitudes to the cavity state is zero in the former case,
 while that to the bright states is zero in the latter case, as will be discussed below.... remove this... state it later somewhere?
 }
 Later, we focus on the optical absorption 
 as the main response function of the system.

\subsection{Fano resonances}
\label{sec:fanosol}

\subsubsection{Optical absorption vs. inelastic electron scattering}
\label{sec:avsm}


For a generic process that can excite the system from 
$\ket{vac}$ to an eigenstate $\ket{\Psi(\w)}$,
the excitation probability \emph{normalised} with
the bare excitation probability of the continuum state
$\ket{\sttt_{0}(\w)}$
is given by the Fano formula
~\cite{Fano61,ZebFano}.
Let's assume $\hat T$ to be the operator 
that
describes the excitation of molecules in 
an inelastic electron scattering event.
  Obtaining this spectrum in experiments on microcavities is likely to be the easiest when a gas of molecules is confined between the cavity mirrors or flows through them.
 In typical organic microcavities with active molecules in a solid film of thickness $\sim100nm$ sandwiched between two plane mirrors,
 the exposed region of the film along its thickness can be probed.
We can consider the excitation of the dark 
molecular states (eigenstates of $\h_D$ in Eq.~\ref{eq:hd}) 
as a background and focus on the bright band.
The normalised probability for this process is given by~\cite{Fano61,ZebFano},
\begin{align}
\nonumber
\mathcal{M}(\w) &\equiv 
\frac{|\braket{\Psi(\w)|\hat T|vac}|^2}{|\braket{\sttt_{0}(\w)|\hat T|vac}|^2},\\
\label{eq:Mw}
&=
\frac{(q+\epsilon)^2}{1+\epsilon^2},
\end{align}
where 
\com{ the reduced energy $\epsilon$ and 
the asymmetry ``parameter'' $q$
are given by~\cite{ZebFano}
}
\begin{align}
\nonumber
\epsilon &= \frac{\w-\w_c-F(\w)}{\pi |V(\w)|^2},\\
\label{eq:fe}
&=\sqrt{\frac{2}{\pi}}
\frac{\sigma}{\wrr^2} e^{\w^2/2\sigma^2} (\w-\w_c) - \erfi(\w/\sqrt{2}\sigma),
\end{align}
and the ``asymmetry parameter'' $q$ is given by
\begin{align}
\nonumber
q &= \frac{1}{\pi V(\w)} 
\frac{\braket{\Phi(\w)|\hat T|vac}}
{\braket{\sttt_{0}(\w)|\hat T|vac}},\\ \nonumber
&= \frac{1}{\pi V(\w) }
p.v.\int{\frac{V(\w')d\w'}{\w-\w'}}
,\\
\label{eq:fqe2}
&= \erfi(\w/2\sigma),
\end{align}
where we 
assumed the bare transition matrix element 
${\braket{\sttt_{0}(\w)|\hat T|vac}}$
is constant over the energy range of interest
and used ${\braket{1_P|\hat T|vac}=0}$ as 
the electron scattering would not create cavity photons.

\com{
In the weak coupling regime $\epsilon$ is a simple monotonic function of 
$\w$ while the ``asymmetry parameter'' $q$ is a constant
at relevant energies ${\w\lesssim\wrr\ll\sigma}$~\cite{Fano61,ZebFano}.
Since ${\braket{1_P|\hat T|vac}=0}$ in our case,
q is ${\mathcal{O}(\w/\sigma)\sim 0}$ in the weak coupling regime.
}

For the optical absorption in our hybrid system, however,
the bare transition amplitude to the bright continuum is zero, i.e.,
${\braket{\sttt_{0}(\w)|\ad|vac} = 0}$,
where
$\ad$ is the transition operator for this process.
We thus use the \emph{unnormalised} probability
given by,
\begin{align}
\nonumber
\mathcal{A}(\w) &\equiv  
|\braket{\Psi(\w)|\ad|vac}|^2,\\ 
\label{eq:Aw}
&= \frac{\wrr^2 P(\w)}{[\w-\w_c-F(\w)]^2 + \pi^2 \wrr^4 P(\w)^2},
\end{align}
which is simply the photon spectral function.
In the context of the weak coupling Fano resoance,
the vanishing of ${\braket{\sttt_{0}(\w)|\ad|vac}}$ also means 
that ${q=\infty}$,
so the lineshape in the weak coupling regime is 
a Lorentzian whose width is given by the interaction strength~\cite{Fano61,ZebFano}.
The above result is exact as long as there is no 
cavity leakage or non-radiative emitter losses.
We will include these using Green's function formalism
later.

\begin{figure}
\centering
\includegraphics[width=1\linewidth]{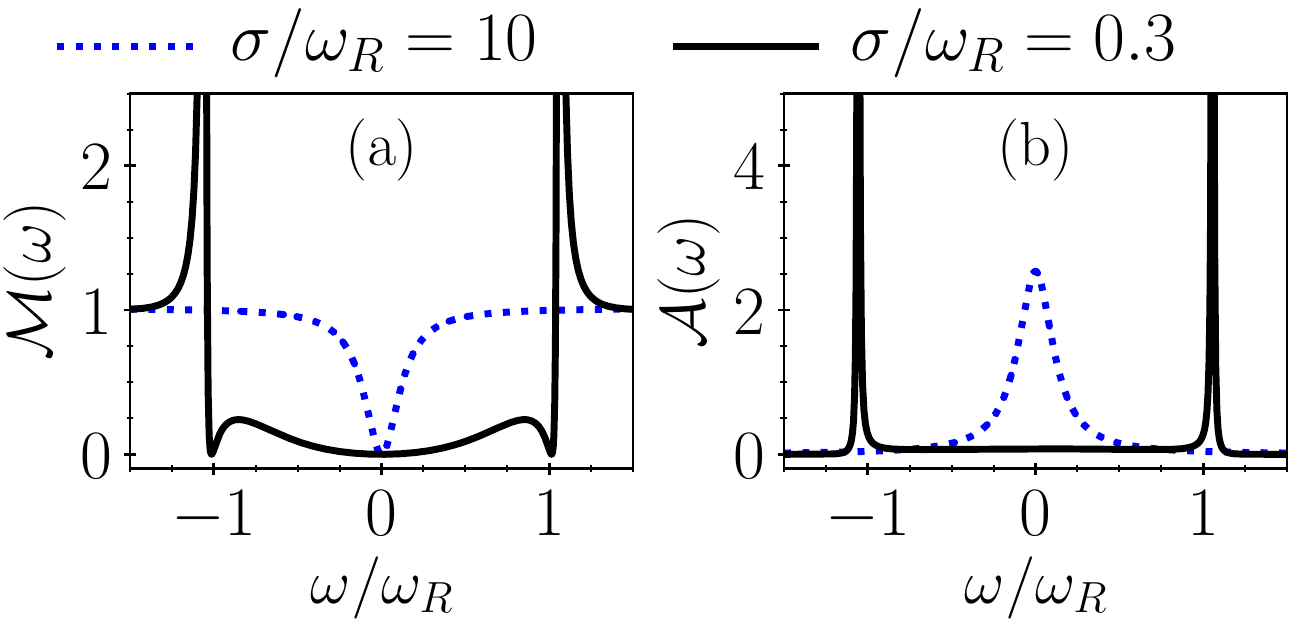}
\caption{Fano resonances in (a) $\mathcal{M}(\w)$ and (b) $\mathcal{A}(\w)$
at $\sigma/\wrr=10,0.3$ and $\w_c=0$.
Blue dotted lines are standard weak coupling Fano resonances 
(with asymmetry parameter ${q\simeq 0,\infty}$)
while 
the black lines are their strong coupling~\cite{ZebFano} counterparts.
}
\label{fig:avsm}
\end{figure}

\com{
In the classical coupled oscillators case~\cite{},
this would be like coupling the probe to only one of the oscillators but not both.
}

\subsubsection{Fano resonances in the weak and strong coupling regimes}
\label{sec:fres}

Both $\mathcal{M}(\w)$ and $\mathcal{A}(\w)$ give information 
about the eigenstates of $\hex$.
However,
their lineshapes are completely different due to the fact that 
they either excite the molecules or the cavity but not the both
(as ${\braket{\sttt_{0}(\w)|\ad|vac}=0=\braket{1_P|\hat T|vac}}$).
Figure~\ref{fig:avsm} shows
$\mathcal{M}(\w)$ and $\mathcal{A}(\w)$
at ${\sigma/\wrr=10,0.3}$
and ${\w_c=0}$.
At ${\sigma/\wrr=10}$ (blue dotted curves), 
the system is in the weak coupling regime~\cite{ZebFano}
where we see the famous characteristic profiles around ${\wt_c=0}$
corresponding to ${q\simeq0,\infty}$
---
inverted Lorentzian dip in ${\mathcal{M}(\w)}$
and a Lorentzian peak in ${\mathcal{A}(\w)}$.
In usual terms,
the dip arises due to a destructive interference between two transition paths, 
one to 
the bare bright states $\sket{\w}$
and the other to the modified (and broadened) discrete state 
$\dsket$
(that now has
components of the bright band as well).
${\mathcal{M}(\w)}$ drops to zero at the dip, 
indicating a complete destructive interference.
For the same eigenstates, ${\mathcal{A}(\w)}$ 
behaves differently because
it only excites the photon state.

\com{Note that we do not have any homogeneous broadening
as the intrinsic linewidths of the cavity and the emitters
are not considered here.
--- no need to ring bells and distract??
}

At $\sigma/\wrr<1$,
two new resonances appear 
at $\wt_{LP}$ and $\wt_{UP}$
where
the phase angle $\Delta(\w)$ jumps discontinuously~\cite{ZebFano},
which will now be seen in the two spectra.
The black curves in Fig.~\ref{fig:avsm} show
${\mathcal{M}(\w)}$ and ${\mathcal{A}(\w)}$ at
$\sigma/\wrr=0.3$.
We see sharp peaks around ${\w/\wrr=\pm1}$ in either spectrum,
but the peak profiles of ${\mathcal{M}(\w)}$ and ${\mathcal{A}(\w)}$ are still very different.
While ${\mathcal{A}(\w)}$ is not a superposition of two Lorentzians,
as two (homogeneously) broadened polariton peaks would produce in case of identical emitters,
it looks quite similar.
However,
${\mathcal{M}(\w)}$ shows dips adjacent to these peaks
where ${\mathcal{M}(\w)}$ vanishes,
clearly establishing 
the difference between the eigenstates of our model from 
broadened polariton states, which
will be discussed later in sec.~\ref{sec:homo}.
While going from the weak to the strong coupling regime,
a residual of original resonance still exists.
At $\sigma/\wrr=0.3$, it is too small to be visible
 in ${\mathcal{A}(\w)}$ 
but shows its signatures in ${\mathcal{M}(\w)}$
that always vanishes at ${\w=\wt_c=0}$.

\com{
In the limit $\sigma/\wrr\lll1$,
these new strong coupling peaks become ultra sharp 
and
move to
the exact locations of $\wt_{LP},\wt_{UP}$, 
which match the polariton energies given by Eq.~\ref{eq:EUP},\ref{eq:ELP}
in this limit, thus
describing
the discrete polaritons states of finite $W$ case.
}

\com{Due to this correspondence between these
eigenstates $\sigma/\wrr\lesssim1$ 
and $\sigma/\wrr\lll1$,
the former will be called polariton's Fano resonances in this article.
}

\begin{figure}
\centering
\includegraphics[width=1\linewidth]{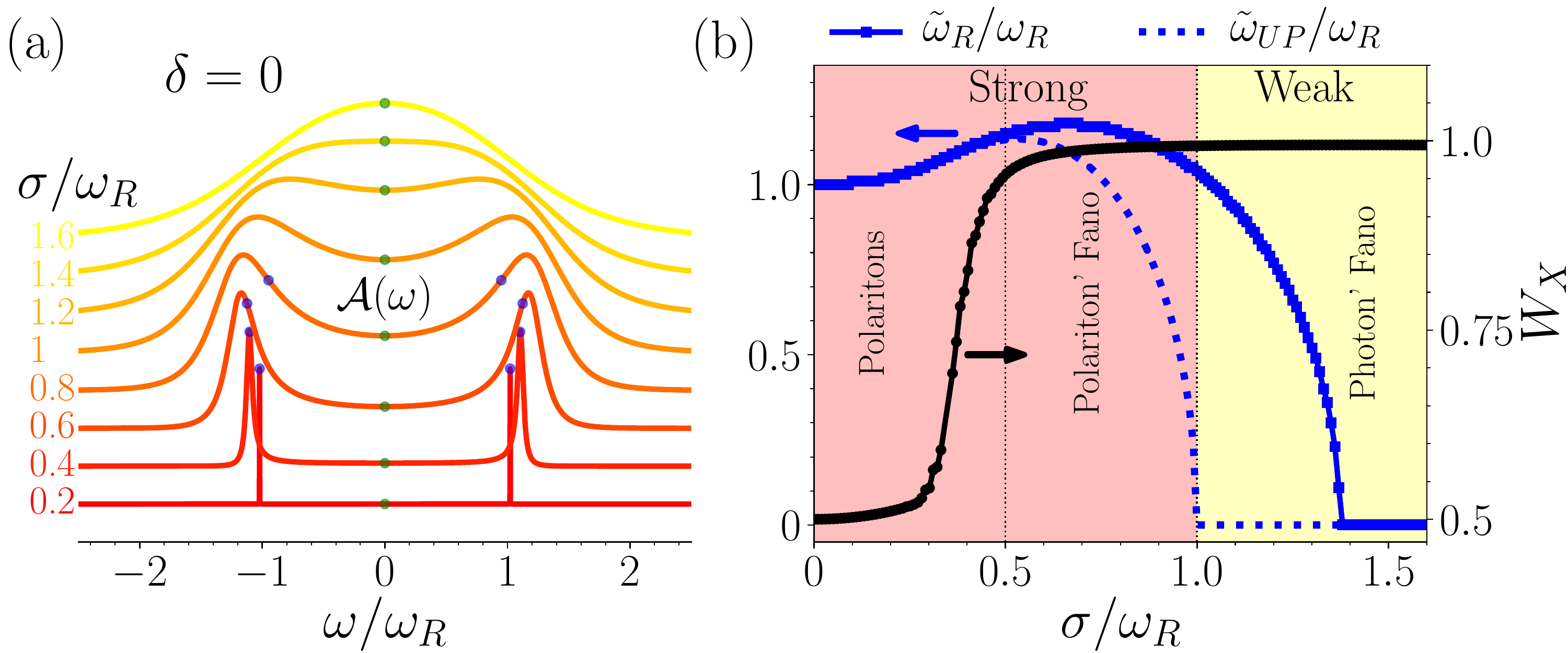}
\caption{The evolution of the Fano resonances with disorder strength $\sigma$ for a resonant cavity mode ($\delta=0$). 
(a) The spectral weight $\mathcal{A}(\w)$ at
${\sigma/\wrr\in[0.2,1.6]}$ shown red to yellow, values labelled on the left.
As $\sigma$ increases, the polariton peaks broaden due to interactions and become polaritons' Fano resonances,
which then merge together and give the photon's Fano resonance. 
The blue and green dots show the actual locations $\wt_{UP},\wt_{LP},\wt_c$ of the resonances as determined by the jumps in the phase $\Delta(\w)$.
(b) The Rabi splitting $\tilde \omega_R$ 
or the location of the upper polariton peak in ${\mathcal{A}(\w)}$
and $\wt_{UP}$ as a function of $\sigma$ again at ${\delta=0}$ as in (a). 
The emitter weight $W_X$ in the state at the absorption peak is also shown, see the scale on the right. 
A finite splitting (${\tilde \omega_R>0}$) 
exists even in the weak coupling regime ${\sigma/\wrr\geq1}$~\cite{ZebFano}.
Furthermore, even the strong coupling does not guarantee the \emph{hybrid} excitations, ${W_X\simeq 1}$ coexisting with ${\tilde \omega_R\sim\wrr}$ in the middle region.
\com{
$W_X$ increases from $1/2$ to $\simeq 1$ around $\sigma/\wrr\simeq 0.5$, whereas $\tilde \omega_R$ stays close to $\wrr$ until $\sigma/\wrr\simeq 1$.
So, a finite Rabi splitting does not guarantee the \emph{hybrid} excitations.
Three regions are labelled according to $\tilde \omega_R$ and $W_X$. 
}
}
\label{fig:fano}
\end{figure}

\com{
Herrera and spano, and Dong, and other papers on organics that deal with vibrations etc... how to cite them...? do I need to discuss them at length? just because referee like me to compare my results with them...?
}

\section{Effect of energetic disorder on the eigenstates} 
\label{sec:absorption}

\subsection{Rabi splitting and polaritons}
\label{sec:rabi}

We now focus on the optical absorption and 
see how the Rabi splitting $\tilde \omega_R$ observed in the absorption spectrum
depends on the energetic disorder $\sigma$,
and wether a finite splitting necessarily means the formation of polaritons or the strong matter-light coupling.
This is an important question as the optical spectrum
can exhibit a similar two-peak structure 
even when the underlying quantum states are entirely different,
as explained below.
In the absence of energetic disorder,
$\hex$ has \emph{only two polariton states} given in sec.~\ref{sec:identical},
which are observed in experimental optical spectra as two sharp peaks at $\w_{UP},\w_{LP}$ (Eq.~\ref{eq:EUP},\ref{eq:ELP})
with a non-zero linewidth due to homogeneous broadening $\gamma$ and 
cavity leakage $\kappa$ ($\kappa,\gamma\ll\wrr$).
However, in the presence of energetic disorder, we have \emph{a continuum of states}
$\evket$
that can also produce two peaks in the absorption just like the polaritons in the above case.
In the following, we first present our results of optical spectrum which pertains to the states $\evket$.
We will discuss the differences from the other case described above in sec.~\ref{sec:homo}.

\com{
Ignoring the intrinsic linewidths of the cavity and the emitters, i.e.,
${\kappa=0=\gamma}$,
the optical absorption, given by Eq.~\ref{eq:Aw},
still bears a finite broadening as already seen in Fig.
}

Figure~\ref{fig:fano}(a) shows the evolution of 
the photon spectral weight
$\mathcal{A}(\w)$
with $\sigma$ for ${\sigma/\wrr\in[0.2,1.6]}$ (shown red to yellow)
for a resonant cavity, ${\delta=\w_c-\w_0=0}$.
The roots of Eq.~\ref{eq:roots}
are also shown as blue ($\wt_{UP}, \wt_{LP}$) 
and green (${\wt_c=0}$) dots to compare them with the peaks positions.
At small $\sigma$, we have two sharp peaks 
for the upper and the lower polaritons
which broaden and shift as $\sigma$ increases,
and finally merge together into a single  broad central peak.
At ${\sigma/\wrr=0.2}$, the peaks are ultra sharp 
and their locations agree to $\wt_{UP}, \wt_{LP}$
(and the clean case $\pm\wrr$, Eqs.~\ref{eq:EUP},\ref{eq:ELP}).
This can be understood by ignoring the bright states 
away from $\w\simeq0$ as they do not have significant effect
on the polaritonic eigenstates formed by the coupling between
$\ket{1_P}$ and $\ket{\sttt_0(0)}$.
However,
at higher ${\sigma}$, still in the strong coupling case,
the peaks are shifted slightly outwards from $\wt_{UP}, \wt_{LP}$.
This can also be explained as 
arising from the coupling of $\ket{1_P}$ with the detuned bright states
on either side of the ${\w=0}$.
Consider the bright states at ${\w\gtrsim0}$. It will push both lower and upper polaritons
upwards by an amount that would depend on their detuning from the polaritons, which is ${\wrr-\w}$ for the upper polariton but ${\wrr+\w}$
for the lower polariton.
So the effect on the upper polariton will be stronger.
Similarly, the bright states at ${\w\lesssim0}$ that pull down the polaritons 
will affect the lower polariton more strongly,
leading to a net upward shift of the upper polariton and a net downward shift of the lower polariton.
The bright states within the two polaritonic states will have this effect.
At ${\sigma/\wrr\lesssim1}$, 
the effective collective coupling of these states becomes weaker
and 
comparible to the bright states 
outside this polaritons window,
so
all these states obtain significant photon component
leading to excessively broad absorption peaks.
At $\sigma/\wrr\ge1$, $\wt_{UP}, \wt_{LP}$ do not exist and 
we are in the weak coupling regime with a single root $\wt_c=0$,
but, as we can see, $\mathcal{A}(\w)$ still keeps its two-peak structure at $\sigma/\wrr=1,1.2$.

Thus, as $\sigma$ increases, the interactions 
transform
the polaritons into two polariton's Fano resonances,
which are eventually replaced by 
a single photon Fano resonance at $\sigma/\wrr\ge1$.
For clarity, we call it a polariton's Fano resonance when the hybrid polaritonic character of the eigenstate is compromised due to Fano broadening.
We can think of the polariton's Fano resonances as arising from the interaction of polaritons (formed by cavity mode and bright states that are resonant or more strongly coupled to it)
and off-resonant or less strongly coupled bright states,
as illustrated in Fig.~\ref{fig:cartoon}(c-d) and also
discussed in sec.~\ref{sec:polout}. 
However, it is worth noting that
this is not strictly the same as Fano resonances arising from a weak coupling between two discrete states and a continuum~\cite{Fano61}.
At large enough $\sigma$, 
there are not enough states or coupling strength to create the strong coupling resonances, 
so we obtain a single photon Fano resonance in this weak coupling regime.

It is also interesting to see that
at $\w=\wt_{UP}, \wt_{LP},\wt_c$,
 the expression for $\mathcal{A}(\w)$ in Eq.~\ref{eq:Aw}
 can be simplified (as the first term in the denominator vanishes)
 to obtain
${\wrr\mathcal{A}(\w)=\sqrt{2/\pi^3}\sigma/\wrr\,\exp(\w^2/2\sigma^2) }$,
so that the size of the central peak is always proportional to $\sigma$.
Since the spectral function is normalised, increasing $\sigma$
thus decreases the linewidth of the central peak at $\sigma/\wrr>1$.
This gives the correct picture in case of $\sigma/\wrr\gg1$
where the coupling becomes too weak and the cavity photon state
is only slightly perturbed by the interaction with the bright continuum.

In Fig.~\ref{fig:fano}(b),
the Rabi splitting $\tilde{\w}_R$ 
or the position of the upper polariton peak in 
$\mathcal{A}(\w)$,
and the actual location of the related resonance $\wt_{UP}$,
are shown as a function of $\sigma$
again at ${\delta=0}$. 
We see that, as $\sigma$ increases,
$\tilde{\w}_R$ slightly increases at small $\sigma$ and then decreases sharply to zero at large $\sigma$
where the two peaks in Fig.~\ref{fig:fano}(a) merge together.
$\wt_{UP}$ also shows a similar behaviour but slightly quicker
leading to an interesting situation
where 
$\mathcal{A}(\w)$ has two peaks with
$\tilde{\w}_R>0$ even in the weak coupling regime, 
$\sigma/\wrr\geq 1$.
In the strong coupling case,
these peaks should exist
because
the first term in the denominator of Eq.~\ref{eq:Aw}
becomes non-monotonic~\cite{ZebFano})
(vanishes at multiple locations)
but 
here it is sustained because
its sum with the second term still bears a bimodal structure.

Thus we find important implications for the experiments on disordered systems: 
the Rabi splitting, or the anticrossing in the optical absorption or transmission spectrum alone cannot guarantee
 the formation of polaritons or the \emph{strong coupling}.
Besides,
strictly speaking, these two terms should not be considered 
synonymous in case of disordered systems as there is a wide range of $\sigma/\wrr$ 
in the strong coupling regime 
where even the eigenstates containing the largest photon spectral weight are still 
far from truly hybrid polaritons.
This will be further explained in the following section.

\subsection{Inhomogeneous vs homogeneous broadening}
\label{sec:homo}

Let's see how the effects of an
\emph{inhomogeneous} broadening or energetic disorder $\sigma$
are different from that of a 
\emph{homogeneous} broadening
or the intrinsic linewidth $\gamma$ of the emitter states.
We can compare the spectrum 
in Fig.~\ref{fig:fano}(a) 
to that of a system with no energetic disorder but a large 
\emph{homogeneous} broadening, 
e.g., atoms in a microcavity as in Ref.~\cite{ThompsonPRL92} (see Fig.~2 there).
While the spectra look similar, their physical interpretation 
depends on the nature
of the quantum states
of the system, which is different in the two cases.
The homogeneous broadening of the emitter states $\gamma$
broadens the polaritonic eigenstates or polaritons
around their expected energy ($\pm\wrr$ at $\delta=0$)
and keep their polaritonic character 
intact.
In contrast,
the inhomogeneous broadening $\sigma$
(that creates the continuum of the bright emitter states in the first place)
broadens the photon spectral function only 
without broadening
the eigenstates,
making the polariton states at the absorption peaks
 more excitonic or emitter-like
as $\sigma$ increases.
That is, the polaritonic character of the eigenstates is distributed over 
larger number of eigenstates, making even
the most polaritonic state nearly excitonic 
when the Fano broadening spans 
a large number of emitter-states in the bright continuum.

This is best illustrated by 
considering a specific density of continuum states
and 
numerically evaluating 
the emitter weight of the most polaritonic eigenstate.
For example,
taking one bright state every ${\Delta\w=0.003\wrr}$,
the emitter weight $W_X$ of 
the states at the absorption peaks 
is plotted in Fig.~\ref{fig:fano}(b) [black curve] as a function of $\sigma$ at $\delta=0$.
We see that $W_X\sim 1/2$ (polaritons) at $\sigma\lesssim 0.3$ where
the Fano broadening is still smaller than the ``width'' $\Delta\w$ of a single eigenstate,
but quickly approaches unity around $\sigma\simeq 0.5$
where the Fano broadening 
distributes the
photon spectral weight over many eigenstates.
Thus polaritons lose their meanings and the peaks 
in the optical absorption simply correspond 
to strongly absorbing emitter states.
As the figure shows, 
this happens at $\sigma/\wrr$ as low as $0.5$
with $\tilde{\w}_R\simeq \wrr$, i.e., well below
the threshold for the weak coupling $\sigma/\wrr=1$.

This raises another question that ought to be discussed here. 
What does $\Delta\w$ corresponds to in a real system?
An intuitive answer to this question is that 
it plays the role of the intrinsic linewidth or the \emph{homogeneous} broadening $\gamma$ of the emitter states
(i.e., ${\Delta\w\simeq\gamma}$).
Increasing the homogeneous broadening $\gamma$ for
the emitter states should extend the bright band on either side by ${\gamma/2}$
but reduce the number of such homogeneously broadened 
bright states in the band (${\simeq W/\gamma}$). 
This should enhance the polaritonic character of the states at the cost of 
their broadening.
In the extreme case ${\gamma\to W}$,
 the bright band will become a homogeneously broadened bright state that would produce two homogeneously broadened polaritons as
observed in Ref.~\cite{ThompsonPRL92}.

\begin{figure}
\centering
\includegraphics[width=1\linewidth]{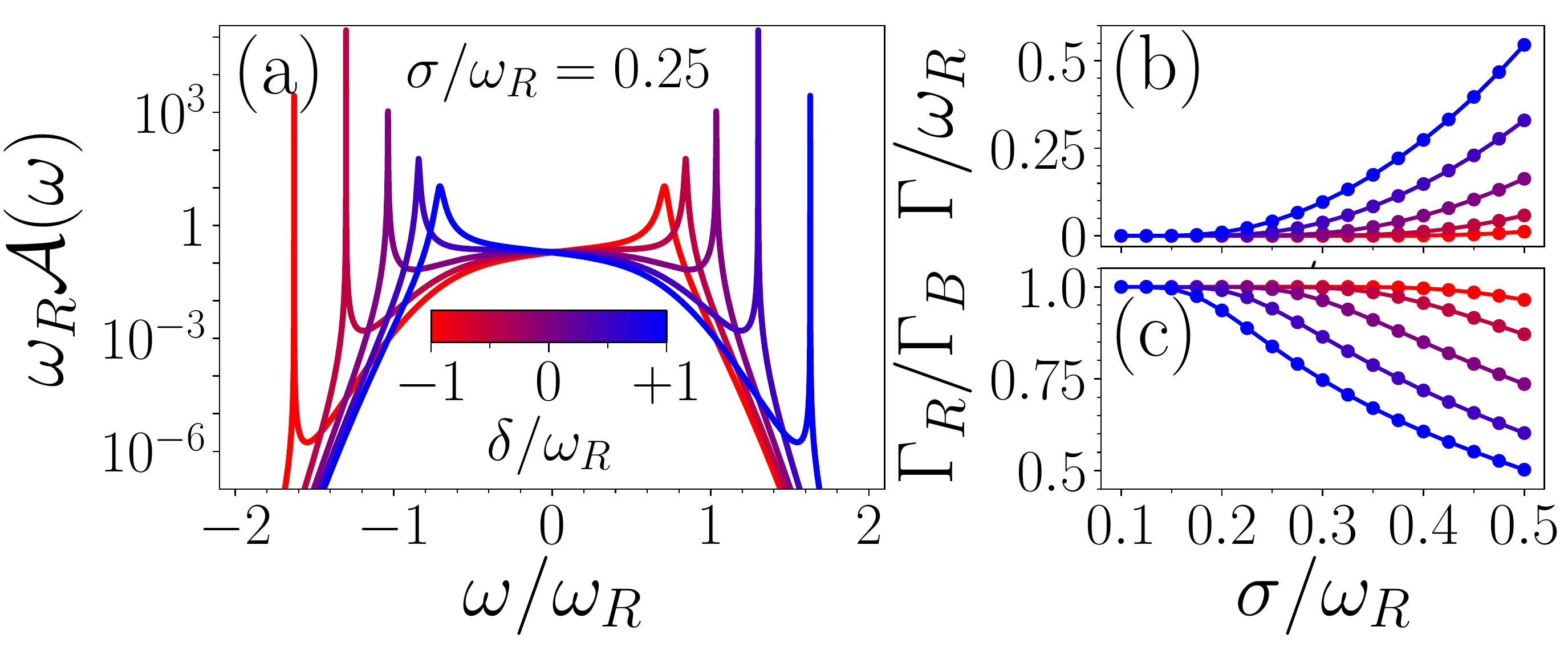}
\caption{Fano broadening as a function of detuning $\delta$
and disorder strength $\sigma$.
(a) Optical absorption at $\delta/\wrr\in[-1,1]$
shown red to blue
at $\sigma/\wrr=0.25$.
The closer a peak is to the centre of the bright band, the broader it gets. 
(b,c) The linewidth $\Gamma$
and the ratio of the half widths $\Gamma_R/\Gamma_B$
 of the lower polariton peak as a function of $\sigma$ at the same parameters as in (a).
$\Gamma$ increases almost quadratically with $\sigma$.
 The effect is stronger for the positive detuning where the lower polariton peak is closer to the bright band centre.
The deviation of $\Gamma_R/\Gamma_B$ from unity 
also follows the trend of $\Gamma$.
}
\label{fig:vsd}
\end{figure}

\subsection{Fano broadening: shape and width of polaritonic peaks}
\label{sec:fbroad}

Let's focus on the Fano broadening and the lineshape asymmetry
in the strong coupling regime even when the corresponding weak coupling case exhibits a symmetric Lorentzian profile.
We find that
the Fano broadening appears even at fairly small $\sigma$. 
Figure~\ref{fig:vsd}(a)
shows the scaled spectral weight ${\wrr\mathcal{A}(\w)}$ 
at ${\sigma/\wrr=0.25}$
and ${\delta/\wrr\in[-1,+1]}$.
We see that as a polariton peak gets near the
bright emitter band,
its linewidth increases and it transforms into a Fano resonance.
This also accompanies a slight shift in the peak position 
compared to the clean case ${\sigma\to0}$,
Eqs.~\ref{eq:EUP},\ref{eq:ELP}, 
and the actual locations of the resonance ${\wt_{UP},\wt_{LP}}$,
as discussed before for ${\delta=0}$. 
Considering the peak corresponding to the lower polariton,
Fig.~\ref{fig:vsd}(b,c) show
its linewidth $\Gamma$ and a simple measure of the asymmetry of its lineshape, the ratio of the half linewidths of the low ($\Gamma_R$) and the high ($\Gamma_B$) energy side of the peak,
as functions of $\sigma$ for the same set of $\delta$ values 
as in Fig.~\ref{fig:vsd}(a).
We see that $\Gamma$ increases with $\sigma$ much more strongly at large positive detuning (bluish curves) where the lower polariton 
is closer to the bright emitter band and
would be more matter-like even at $\sigma=0$.
The lineshape asymmetry shown in Fig.~\ref{fig:vsd}(c), the deviation of ${\Gamma_R/\Gamma_B}$ from $1$, also shows the same trend.

We will now consider a small homogeneous broadening 
and treat it
using the standard Green's function formalism.
To treat both homogeneous and inhomogeneous broadening of 
the emitter states on equal footing,
new methods need to be developed,
which is clearly beyond the scope of the present work.

\begin{figure}
\centering
\includegraphics[width=1\linewidth]{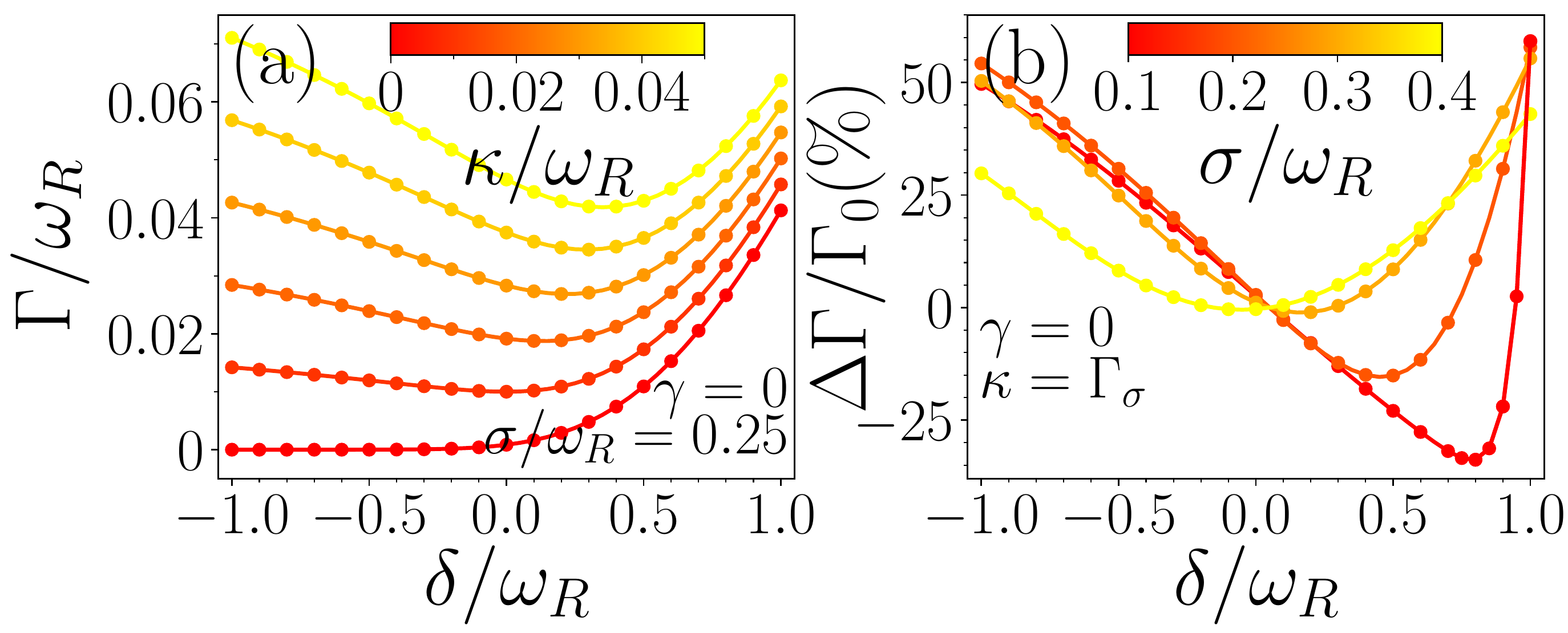}
\caption{
\com{Fano broadening as a function of disorder strength $\sigma$ and detuning $\delta$.
(a) Fano resonance at $\delta/\wrr\in[-1,1]$
shown red to blue
at $\sigma/\wrr=0.25$ and $\kappa=0=\gamma$.
The closer a peak is to the centre of the bright band, the broader it gets. 
(b,c) The linewidth $\Gamma$
and the ratio of the half widths $\Gamma_R/\Gamma_B$
 of the lower polariton peak as a function of $\sigma$ at the same parameters as in (a).
$\Gamma$ increases almost quadratically with $\sigma$.
 The effect is stronger for the positive detuning where the lower polariton peak is closer to the bright band centre.
The deviation of $\Gamma_R/\Gamma_B$ from unity 
also follows the trend of $\Gamma$.
}
Total broadening as a function of disorder strength $\sigma$ and detuning $\delta$ in presence of a small homogeneous broadening $\kappa$ of the cavity state.
(a) $\Gamma$ as a function of $\delta$ at $\kappa/\wrr\in[0,0.05]$ shown red to yellow.
$\Gamma \to \kappa$ at large negative $\delta$
while ${\Gamma \to \Gamma_\sigma=\Gamma(\sigma,\gamma=0=\kappa,\delta/\wrr=+1)}$ at large positive $\delta$ [the blue curve in Fig.~\ref{fig:vsd}(b)].
(b) The percentage change in $\Gamma$ as $\delta$ changes
from zero, at $\sigma/\wrr\in[0.1,0.4]$, $\kappa=\Gamma_\sigma$, $\gamma=0$.
}
\label{fig:vsdk}
\end{figure}

\section{Optical absorption in the presence of losses}
\label{sec:greenlosses}

So far, we have not considered the losses,
a finite cavity leakage and non-radiative emitter decay
that amount to homogeneous broadening of the bare cavity and emitter states.
To see the effect of these losses on the spectral function $\mathcal{A}(\w)$
or optical absorption,
we need the photon Green's function $G(\w)$
as ${\mathcal{A}(\w)=-\Im G(\w)}$, 
which will now be used instead of
Eq.~\ref{eq:Aw}.
Here $\Im$ stands for the imaginary part.
Due to the absence of the coupling between the emitters or their bright states, 
$G(\w)$ can be easily calculated from the resolvent of $\hex$ (Appendix.~\ref{asec:green}),
to obtain
\begin{align}
\label{eq:Gw}
G(\w)&=\left[\w-\w_c+i \kappa - \Sigma(\w+i \gamma) \right]^{-1},\\
\label{eq:Sigz}
\Sigma(z)
&= \wrr^2 \int d\w'   \frac{P(\w')}{z - \w'},
\end{align}
where $\Sigma$ is the self energy and
$\kappa,\gamma$ are cavity leakage and non-radiative emitter relaxation rates.
Computing the self energy $\Sigma(z)$
for the Gaussian $P(\w)$ readily gives
\begin{align}
\Sigma(z)
\label{eq:Sigzg}
 &= \pi \wrr^2 P(z) \left[\erfi\left(\frac{z}{\sqrt{2} \sigma }\right)-i\right],
\end{align}
which can now be used with Eq.~\ref{eq:Gw}
to explore the optical absorption including the losses.

\subsection{Lineshape and linewidth}
\label{sec:lineshape}

\com{
First, let's focus on the Fano broadening and the lineshape asymmetry
in the strong coupling regime even when the corresponding weak coupling case exhibits a symmetric Lorentzian profile.
We find that
the Fano broadening appears even at fairly small $\sigma$. 
Figure~\ref{fig:vsdk}(a)
shows the scaled spectral weight ${-\wrr\Im G(\w)}$ at ${\sigma/\wrr=0.25}$, ${\delta/\wrr\in[-1,+1]}$, and ${\kappa=0=\gamma}$ again.
We see that as a polariton peak gets near the
bright emitter band,
its linewidth increases and it transforms into a Fano resonance.
This also accompanies a slight shift in the peak position 
compared to the clean case $\sigma\to0$,
Eqs.~\ref{eq:EUP},\ref{eq:ELP}, 
as discussed before. 
Considering the peak corresponding to the lower polariton,
Fig.~\ref{fig:vsdk}(b,c) show
its linewidth $\Gamma$ and a simple measure of the asymmetry of its lineshape, the ratio of the half linewidths of the low ($\Gamma_R$) and the high ($\Gamma_B$) energy side of the peak,
as functions of $\sigma$ for the same set of $\delta$ values and $\kappa,\gamma$ as in Fig.~\ref{fig:vsdk}(a).
We see that $\Gamma$ increases with $\sigma$ much more strongly at large positive detuning (bluish curves) where the lower polariton 
is closer to the bright emitter band and
would be more matter-like even at $\sigma=0$.
The lineshape asymmetry shown in Fig.~\ref{fig:vsdk}(c), the deviation of ${\Gamma_R/\Gamma_B}$ from $1$, also shows the same trend.
}

Let's now include the effect of cavity and emitter losses on the total broadening of absorption peaks.
In systems where $\kappa\gg\gamma$, e.g., 
organic microcavities~\cite{Cwik16,Dimitrov16},
we can ignore $\gamma$ to understand how $\kappa$
changes the picture discussed so far.
Figure~\ref{fig:vsdk}(a) shows $\Gamma$ as a function of $\delta$ at $\sigma/\wrr=0.25$, $\gamma=0$ and $\kappa/\wrr\in[0,0.05]$.
We see that at large negative detuning, $|\delta|\gg \sigma$, where the polariton is away from the band and Fano broadening can be ignored, the effect of $\kappa$ is the strongest due to larger photon fraction.
As $\delta$ increases to zero and positive values, 
the effect of $\kappa$ decreases but that of $\sigma$ increases. 
This leads to an interesting situation where 
$\Gamma$ develops a minimum at an optimum $\delta$.
The minimum gets deeper and more prominent if 
$\Gamma$ at extreme $\delta$ values is large and of similar size. 
Noting that 
at ${\delta/\wrr=-1}$, 
${\Gamma\sim\kappa}$
and at ${\delta/\wrr=+1}$, 
${\Gamma\sim \Gamma_{\sigma}\equiv \Gamma(\sigma, \delta/\wrr=+1,\kappa=0=\gamma)}$ [blue curve in Fig.~\ref{fig:vsd}(b)],
the optimum $\kappa$ to observe 
such a minimum is 
${\kappa=\Gamma_{\sigma}}$.
This is shown in Fig.~\ref{fig:vsdk}(b)
for ${\sigma/\wrr\in[0.1,0.4]}$
where percentage change in $\Gamma$, ${\Delta\Gamma/\Gamma_0=(\Gamma-\Gamma_0)/\Gamma_0}$
is shown as $\delta$ deviates from the resonance, ${\delta=0}$,
where ${\Gamma=\Gamma_0}$, which is used as a reference.
We see that the smaller the $\sigma$, the larger the percentage change in $\Gamma$, and 
the larger the optimum $\delta$. 
It should be observable in typical organic microcavities
where
${\kappa\sim0.05eV}$~\cite{Dimitrov16},
${2\wrr\sim 1eV}$~\cite{TropfAOM18}, and ${\sigma\sim \sqrt{2} \times 0.1eV}$~\cite{bassler81, bassler82,SM}
so $\sigma/\wrr\sim 0.28$,
which gives the minimum at $\delta\sim 0$. 

\begin{figure}
\centering
\includegraphics[width=1\linewidth]{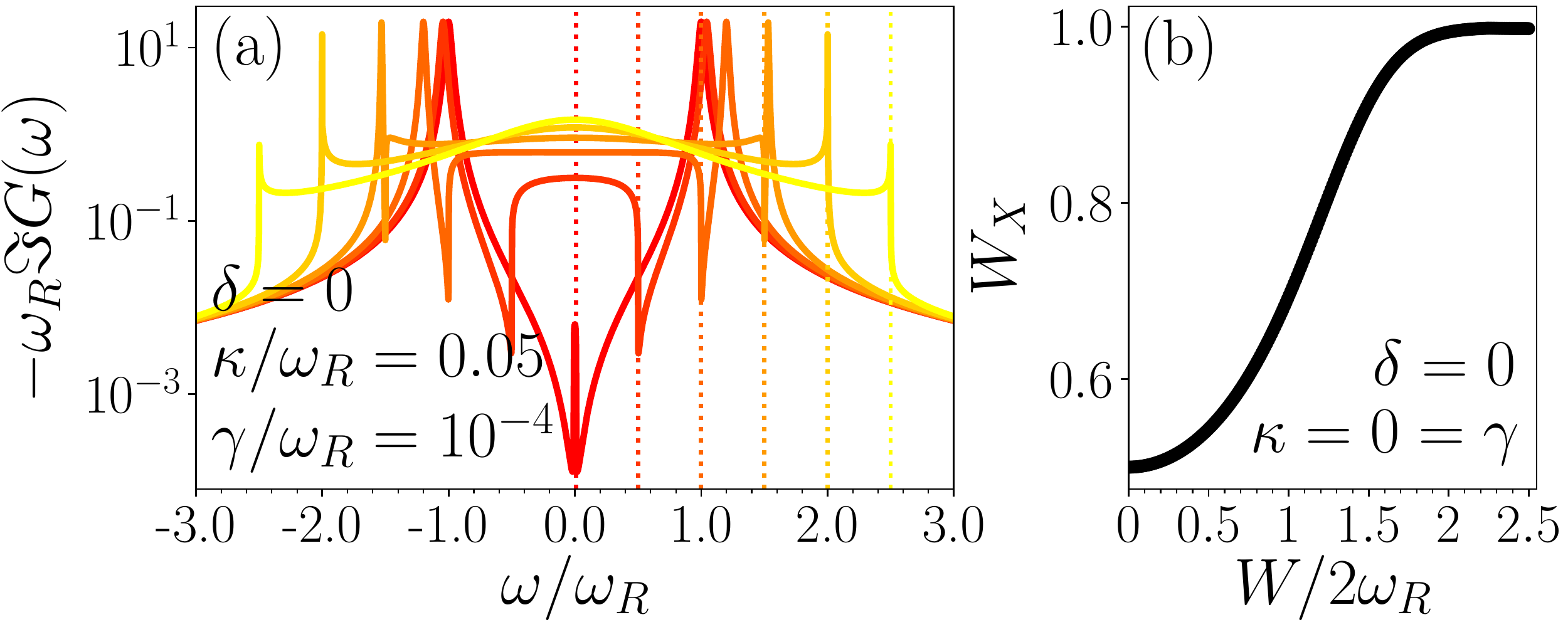}
\caption{Evolution of the Fano resonance with the bandwidth for a flat energy distribution that leads to a flat interaction $V(\w)$. (a) Photon spectral weight $-\wrr\Im G(\w)$ 
at ${\sigma/\wrr=0.01,0.5-2.5}$ (indicated by vertical lines). (b) Emitter weight $W_X$ as a function of the bandwidth $W$.
}
\label{fig:flat}
\end{figure}

\subsection{Dependence on the energy distribution}
\label{sec:distribution}

If the emitter transition energies are distributed more evenly,
the distribution function is flatter in the middle and 
falls more sharply at the edges.
We would expect 
it should result in a sharper transition between polaritons and their Fano resonances as the polariton energy starts overlapping the 
bright states band. 
Taking the extreme case of a flat distribution,
${P(\w)=1/W, -W/2\leq\w\leq W/2, 0}$ otherwise,
we
find that
 the transition 
 actually becomes more gradual. 
This is because in this case all bright states would couple equally strongly to the cavity mode (as ${V(\w)=\wrr/\sqrt{W}=constant}$)
and increasing $W$ would 
push the polaritons away from the band edges at small to intermediate $W$.
In this case, Eq.~\ref{eq:Sigz} gives
\begin{align}
\label{eq:Sigzflat}
\Sigma(z) 
&= \frac{\wrr^2}{W}\ln\left(\frac{z+W/2}{z-W/2}\right).\end{align}
Figure~\ref{fig:flat}(a) shows the scaled spectral weight 
$-\wrr\Im G(\w)$ for the flat distribution (Eq.~\ref{eq:Gw} with Eq.~\ref{eq:Sigzflat})
at ${\delta=0}$, ${\kappa/\wrr=0.05}$, ${\gamma/\wrr=10^{-4}}$ ($\kappa,\gamma$ typical of organic microcavities,~\cite{Cwik16,Dimitrov16})
and ${W/2\wrr=0.01,0.5-2.5}$ in increments of $0.5$,
 marked with dotted vertical lines. 
We see that it has a feature resembling Fano lineshape with a dip
but note the logscale, these structures differ from 
the weak coupling Fano profile indeed.
\com{
as the optical spectrum corresponds to zero bare
transition matrix element for the electronic band states~\cite{ZebFano}.
}
At $W<2\wrr$, the peaks resemble isolated polaritonic states. 
Despite the change in lineshape at $W\ge2\wrr$,
they still remain quite hybrid for a while.
$W_X$ for the lower polariton state
at the peak position at ${\delta=0=\kappa=\gamma}$
is shown in 
Fig.~\ref{fig:flat}(b).
We see that $W_X$ increases from ${1/2}$ to $1$ over a wider $\sigma$ range:
$W_X\simeq 0.7$ only at $W/2\wrr=1$
and it increases to $0.9$ around $W/2\wrr=1.5$.
Compare this to the gaussian energy distribution, where
$W_X$ jumps to $\sim0.95$ around $\sigma/\wrr=0.5$ only.
This sluggish increase in the emitter weight for the flat distribution can be understood as a result of poor correspondence of $\wrr$ itself to the Rabi splitting $\tilde \omega_R$,
because the bright states at the band edge couple equally strongly to the cavity mode and their effective detuning from it
shifts the polariton energy more significantly.
Another difference from the gaussian distribution case
is that the photon Fano resonance appears before the 
polaritons' Fano resonances disappear.
So the notion of the Rabi splitting 
$\tilde \omega_R$ also becomes relatively vague at $W/2\wrr\gtrsim2$ in this case. 
Interestingly, however, for polaritonic peaks, a significant linewidth narrowing occurs
at $W\gtrsim2\wrr$, as can be seen in Fig.~\ref{fig:flat}(a).

\com{
referee: how do we find the location of the resonances....? simple: 
roots of $\w-\w_c-F(\w)=0$... $\Delta$ jumps there and gives the location of the resonances.... 
}

\com{
finite bandwidth... polaritons outside the continuum... 
only a small fraction of the spectral weight is taken by the continuum
if $W/2\ll \wrr$...  Polaritons are discrete states outside the continuum...
Expressions for these polaritonic states... energies $\w_{\pm}$ are roots of $\w_{\pm}-\w_c-F(\w_{\pm})=0$...  

}

\com{
It stays close to $1/2$ (polaritons) at $\sigma/\wrr\lesssim 0.3$
but then sharply increases to $\sim 1$ at $\sigma/\wrr\sim 0.5$,
so that polaritons lose their meanings and the peaks 
in the optical absorption correspond 
simply to strongly absorbing emitter states.
As the figure shows, 
this happens at $\sigma/\wrr$ as low as $0.5$
with $\tilde{\w}_R\simeq \wrr$, i.e., well below
the threshold for the weak coupling $\sigma/\wrr=1$.
}

\section{Summary and conclusions}

We solve the disordered TC model by identifying and
decoupling the exact bright and dark sectors of the emitters' Hilbert space.
Explicit expressions for the bright and the dark states
are presented,
allowing the exact numerical diagonalisation for arbitrarily large systems.
For the emitter states forming a continuum in the thermodynamic limit, 
we obtain Fano's model whose solution has already been around
for a long time.
The Fano resonances in the optical absorption and inelastic electron scattering spectra are presented 
both in the strong and the weak coupling regimes,
where they correspond to polaritons or their Fano resonances and photon's Fano resonance, respectively.
We find that the hybrid light-matter character of the excitations 
could be lost due to strong energetic disorder while 
the optical spectra
still showing the Rabi splitting and 
 anticrossing.
We also explore the effect of cavity and emitter losses on the linewidth and lineshape
by calculating the photon Green's function.
We find that the polariton's linewidth should exhibit a minimum
as a function of the detuning 
when the cavity losses are comparable to the maximum Fano broadening.

\section{Outlook}
We studied the model in the rotating wave approximation that ignores 
the counter rotating terms in the light-matter interaction.
These terms introduce coupling within a given parity subspace 
($N_{ex}$ even or odd) which are usually ignored below the ``ultra-strong'' coupling regime ($\wrr\gtrsim0.3\w_0$)
due to a large energetic difference ($\simeq 2\w_0$) 
between the coupled states.
However, energetic disorder would reduce this energetic difference
and, at large energetic disorder, the highest end of the bright band or the upper polariton in ${N_{ex}=1}$ subspace would come close enough to (or overlap with) the 
lowest end of the bright band or the lower polariton in ${N_{ex}=3}$ subspace
such that their interaction cannot be ignored any longer.
Similarly, the effects of an interaction that does not respect the excitation number parity (coupling to a bath, for instance) on the eigenstates of the system,
would be enhanced even further as it would only require
to reduce the gap ($\simeq\w_0$) between the states of adjacent excitation spaces
(e.g., ${N_{ex}=1}$ and ${N_{ex}=2}$).
Investigation of such effects of the energetic disorder can be carried out in a future study.
Considering a weak coupling between the bright and dark spaces due to dipole-dipole interaction
and 
application of the two spaces in other systems, e.g., those
described by the Anderson impurity model,
and 
exploring the effects of Fano broadening on highly excited states such as
polariton condensate and lasing
 would also be interesting future works.

\begin{acknowledgements}
The author thanks Rukhshanda Naheed for fruitful discussions.
\end{acknowledgements}

\section{Appendix}

\subsection{Calculations of $\dsket$ and $\evket$}
\label{asec:phipsi}

\subsubsection{Eigenstates and energies outside the bright band}

Assuming that $\dsket$ represents a discrete eigenstate
of $\hex$ at energy $\w$
outside the bright continuum,
it can be written as a superposition of the form
\begin{align}
\label{aeq:dsket}
\dsket = \aket + \int {B(\w,\w') \sket{\w'} d\w'},
\end{align}
where 
the integration is over the bright band.
By substituting $\dsket$ into the eigenvalue equation
${\hex \dsket = \w \dsket}$,
projecting it onto $\aket$ and $\sket{\w''}$ in turns,
and
using the orthonormality of the basis states,
 ${\braket{1_P|1_P}=1}$,
${\braket{\mathcal{S}_0(\w')|\mathcal{S}_0(\w'')}=\delta(\w'-\w'')}$,
and 
${\braket{1_P|\mathcal{S}_0(\w')}=0}$,
we obtain the following two coupled equations,
\begin{align}
\w_c + \int V(\w') B(\w,\w') d\w' &= \w,\\
V(\w') + \w' B(\w,\w') &= \w B(\w,\w'),
\end{align}
which can be easily solved as ${\w\neq\w'}$
to get 
\begin{align}
B(\w,\w') &= \frac{V(\w')}{\w-\w'},\\
\label{aeq:roots}
\w &=\w_c + \int{d\w' \frac{V(\w')^2}{\w-\w'}},
\end{align}
where the last equation contains $\w$ 
in the integral on the right side as well
and can be solved for $\w$ 
 numerically when $V(\w)$ is specified.
Substituting the above expression for $B(\w,\w')$
 into the Eq.~\ref{aeq:dsket} above
gives Eq.~\ref{eq:dsket} in the main text,
while Eq.~\ref{aeq:roots} is rewritten as
Eqs.~\ref{eq:roots},\ref{eq:fwint} there.

\com{
Normalisaton of $\dsket$:
$\dsket$ can be normalised easily, the normalisation factor
is given by
\begin{align}
\label{eq:dsnorm}
A(\w) &= 1/\sqrt{1+I(\w)},\\
I(\w) &=\int{\frac{V(\w')^2}{(\w-\w')^2}d\w'},\\
&= \frac{\wrr^2}{\sigma^2} \left[
\frac{\sqrt{2}\w}{\sigma}D(\w/\sqrt{2}\sigma)-1
\right],
\end{align}
where $D$ is the Dawson function
and the last line is obtained by considering 
${V(\w)^2=P(\w)}$ to be a Gaussian distribution of standard deviation $\sigma$.
The physical significance of $A(\w)$
is that its value at the eigenvalue of $\dsket$
gives its transition
dipole matrix element.
}

\subsubsection{Eigenstates inside the bright band}

Assuming $\evket$ is
an eigenstate of $\hex$ at an energy within the bright continuum,
we can again write it as a superposition,
\begin{align}
\evket = \alpha(\w) \aket + \int {\beta(\w,\w') \sket{\w'} d\w'},
\end{align}
where the coefficients $\alpha,\beta$ 
can be written in terms of ${\Delta(\w),F(\w)}$ given in the main text
as (see Ref.~\cite{Fano61} for detailed derivation),
\begin{align}
\alpha(\w) &= \sin\Delta(\w)/\pi V(\w),\\
\beta(\w,\w') &= \alpha(\w)\frac{V(\w')}{\w-\w'} - \cos\Delta(\w)\delta(\w-\w'),
\end{align}
Here, $\delta$ is the Dirac delta function.
Substituting these back, we obtain,
\begin{multline}
\evket = \frac{\sin\Delta(\w)}{\pi V(\w)}
\left( \aket + \int {\frac{V(\w')}{\w-\w'}\sket{\w'} d\w'}\right)\\
- \cos\Delta(\w)\sket{\w},
\end{multline}
which
along with $\dsket$ in Eq.~\ref{eq:dsket} at $\w$ inside the bright band 
 (where it does not represent an eigenstate in general),
gives Eq.~\ref{eq:evket}.

\subsection{Calculation of $G(\w)$}
\label{asec:green}

The photon Green's function
$G(\w)$ can be obtained~\cite{MazhorinPRA22} from the resolvent 
$\mathcal{R}(\w)$ of $\hex$, as follows.
We start from the discrete version $\h_B$ in Eq.~\ref{eq:hb}.
In the single excitation space, 
it can be written in the matrix form as, 
\begin{align*}
\mathcal{H}_{B}
&= \begin{pmatrix}
\w_c&G_{1,K_1}&G_{2,K_2}&...&G_{N,K_N}\\
G_{1,K_1}&\w_1&0&...&0\\
G_{2,K_2}&0&\w_2&...&0\\
... &...&...&...&...\\
G_{N,K_N} &0&0&...&\w_N
\end{pmatrix}.
\end{align*}
Its resolvent $\mathcal{R}(\w)$ is given as
$\mathcal{R}(\w)=(I\w-\mathcal{H}_{B})^{-1}$, where $I$ is identity.
It
can be computed by partitioning $I\w-\mathcal{H}_{B}$ in four blocks
as,
\begin{align*}
I\w-\mathcal{H}_{B} &= 
\begin{pmatrix}
A_{1\times1} & B_{1\times N} \\
C_{N\times 1} & D_{N\times N}
\end{pmatrix},
\end{align*}
to take the inverse, where the subscripts with the matrix elements show
the dimensions of each block.
The matrix element of $\mathcal{R}(\w)$ 
corresponding to the cavity state is
the photon Green's function $G(\w)$, given by~\cite{Lu2002}
${G(\w)=(A-BD^{-1}C)^{-1}}$.
\com{
This gives~\cite{Lu2002},
\begin{align*}
\mathcal{R}(\w)&= 
\begin{pmatrix}
(A-BD^{-1}C)^{-1} & -(A-BD^{-1}C)^{-1}BD^{-1} \\
-(D-CA^{-1}B)^{-1}CA^{-1} & (D-CA^{-1}B)^{-1}
\end{pmatrix},
\end{align*}
whose matrix element corresponding to the cavity state is
the photon Green's function, given by
${G(\w)=(A-BD^{-1}C)^{-1}}$.
}
Since $D$ is diagonal, its inverse can be written directly,
and 
we obtain
\begin{align}
G(\w)&=\left[\w-\w_c+i\kappa- \Sigma(\w) \right]^{-1},\\
\Sigma(\w) &= 
\sum_{n=1}^N \frac{G_{n,K_n}^2}{\w - \w_n+i\gamma},
\end{align}
where we have added the imaginary components $\kappa,\gamma$ to $\w_c,\w_n$
to include a finite cavity leakage and non-radiative emitter relaxation.
Considering the continuous limit 
leads to Eq.~\ref{eq:Gw}.

\label{Bibliography}
\bibliographystyle{apsrev4-1}

%
\end{document}